\documentclass{iopart}

\usepackage{graphicx}

\usepackage{moreverb}
\usepackage{url}

\begin{document}

\title[Data analysis and graphing with R]{Using R for data analysis
  and graphing in an introductory physics laboratory}

\author{Primo\v{z} Peterlin}

\address{University of Ljubljana, Faculty of Medicine, Institute of
  Biophysics,\\
  Lipi\v{c}eva 2, SI-1000 Ljubljana, Slovenia}

\ead{primoz.peterlin@mf.uni-lj.si}

\begin{abstract}
  R is a language and computing environment that has been developed
  for data manipulation, statistical computing, and scientific
  graphing.  In the paper, we demonstrate its use analyzing data
  collected in a few experiments taken from an introductory physics
  laboratory.  The examples include a linear dependence, a non-linear
  dependence, and a histogram.  The merits of R are discussed against
  three options often used for data analysis and graphing: manual
  graphing using grid paper, general purpose spreadsheet software, and
  specialized scientific graphing software.
\end{abstract}


\section{Introduction}
\label{sec:intro}

An experiment is not completed when the experimental data are
collected.  Usually, the data require some processing, an analysis,
and possibly some sort of visualization.  The three most commonly
encountered approaches to data analysis and graphing are
\begin{itemize}
\item manual method using grid paper
\item general purpose spreadsheet software
\item specialized scientific graphing software
\end{itemize}

The manual method involving grid paper and a pencil often still seems
to be the one favoured by the teachers.  Indeed it has a certain
pedagogical merit, for it does not hide anything in black boxes: the
student has to work through all the steps him- or herself.  It is also
the preferred option in the environments where students cannot be
expected to own a personal computer.  However, students find the
method tedious, painstaking and old-fashioned.  In all honesty, their
teachers do not use this method in their own research work.

General purpose spreadsheet software like Microsoft Excel or the
freely available OpenOffice Calc contains most functions required for
data analysis and graphing.  They are commonly available and the
students are generally well-versed in their use.  However,
spreadsheets have their drawbacks.  They are difficult to debug, and
they often require an educated operator in order to produce a visually
satisfying graph without visual distractions.

Finally, there exists specialized software for scientific graphing
such as Origin (formerly MicroCal, Inc., now OriginLab), SigmaPlot
(formerly Jandel Scientific, now Aspire Software International), or
Prism (GraphPad Software).  This is usually the preferred option in a
research laboratory, but with a price of around 1000~EUR per license,
it is usually considered too expensive for a classroom or student
laboratory use.

In this paper, we present what we consider a viable alternative to the
three options considered above.  R \cite{Ihaka:1996,R:Manual} is both
a language and an environment for data manipulation, statistical
computing and scientific graphing.  R is founded on the S language and
environment \cite{Becker:1984} which was developed at Bell
Laboratories (formerly AT\&T, now Lucent Technologies) by John
Chambers, Richard Becker and coworkers.  S strove to provide an
interactive environment for statistics, data manipulation and
graphics.  Throughout the years, S evolved into a powerful
object-oriented language \cite{Becker:1988,Chambers:1998}.  Nowadays,
two descendants exist which build on its legacy: S-Plus, a commercial
package developed by Tibco Software, Inc., and the freely available
GNU~R.

Being part of the GNU project, R is free software, meaning it can not
only be downloaded from the Internet free of charge, but also that its
users are granted access to the source code and are encouraged to
disseminate any derived works.  Its primary source is the
Comprehensive R Archive Network (CRAN),
\url{http://cran.r-project.org/}.  It is available both in source code
or binary form for a few popular computing environments (Microsoft
Windows, Linux, MacOS X).

The potential of R in a classroom environment has already been
examined in various disciplines, \emph{e.g.,} statistics
\cite{Horton:2004}, econometrics \cite{Racine:2002}, and computational
biology \cite{Eglen:2009}.  In this paper, we strive to demonstrate
that it meets the needs in a student physics laboratory as well.  In
the rest of the paper, we first present a quick tour of R, where we
demonstrate its use on three problems from the laboratory course
prepared for first year students of medicine, dental medicine and
veterinary medicine \cite{Bozic:2003}, two of them involving bivariate
data (illustrating a linear and a non-linear dependence) and one
involving univariate data (a histogram of radioactive decay).  We then
discuss the positive and the negative aspects of using R, and finally
present the main conclusions.

\section{A quick tour of R}

\subsection{Starting and quitting R}

R uses a command line interface.  Once we run it, we find ourselves at
the \verb">" prompt.  R contains extensive built-in documentation on
the available functions, which can be obtained through the
\texttt{help()} function or the \texttt{?} shortcut, \emph{i.e.,}
\texttt{?help} or \texttt{help(help)} offers help on the
\texttt{help()} function itself.

We quit R with the \texttt{quit()} command or its shorter version,
\texttt{q()}.  We have the option to save the current session:
\begin{verbatim}
> q()
Save workspace image? [y/n/c]:
\end{verbatim}
If we reply affirmatory (``y''), two files will be saved into the
current working directory: one with the definitions of objects, and
another one with the command line history.  The next time R is run
from the same directory, all the objects from the previous session
will be restored.  This also means it is good practice to use
separate directories for different projects.

\subsection{Reading data}

The data are often provided in the form of an ASCII file, with the
fields either separated by white space or by commas (``comma separated
values'', CSV).  Both forms can be read using the
\texttt{read.table()} function, with the file name as the mandatory
argument, and several optional arguments, \emph{e.g.,} \texttt{dec}
for setting the decimal sign and \texttt{sep} for setting field
separator, which make life easier for R users in continental Europe.
A somewhat more basic function for reading ASCII data is
\texttt{scan()}.

R can also read numerous other formats, \emph{e.g.,} XML files and
Excel spreadsheets (the latter only when running in Microsoft Windows,
using either the RODBC library available on CRAN, or the xlsReadWrite
extension, \url{http://sites.google.com/site/treetron/}), or it can
fetch data from the Web or from an SQL server.

\subsection{Case 1: Linear dependence}

Let us start with a simple example, in which the linear dependence
between the concentration and the conductivity of a dilute electrolyte
is examined.  At 5 different concentrations of electrolyte, the
student takes measurement of the resistance of a beaker filled with
electrolyte to the mark, with a pair of electrodes immersed and
connected into a Wheatstone bridge.  Using one known value for
conductivity, the student calculates the rest of conductivities from
the resistances, plots the points into a scatterplot, fits the best
line through the data points, and, measuring its resistance, finally
determines the unknown concentration of an additional sample.  The
task can be achieved with a simple script in R, shown below.  The
resulting graph is shown in figure~\ref{fig:electrolyte}.

\begin{listing}{1}
concentration <- c(1.e-3, 2.5e-3, 5.e-3, 7.5e-3, 1.e-2)
resistance <- c(5.31, 2.66, 1.40, 0.92, 0.78)

# calculate the conductivity
k <- 1.e-4*resistance[1]
conductivity <- k/resistance

# define the labels on x- and y-axis
xlabel <- expression(paste(italic(c)," [mol/L]"))
ylabel <- expression(paste(sigma," [(",Omega," cm)"^-1,"]"))

# plot the data
plot(concentration, conductivity, xlab = xlabel, ylab = ylabel,
     xlim = c(0, 0.01), ylim = c(0, 7.e-4))
grid(col = "darkgray")

# find the best fit
cond.fit <- lm(conductivity ~ concentration,
               data = data.frame(concentration, conductivity))

# plot the fitted line
abline(cond.fit)
\end{listing}

\begin{figure}
  \centering
  \includegraphics[angle=270,width=0.6\linewidth]{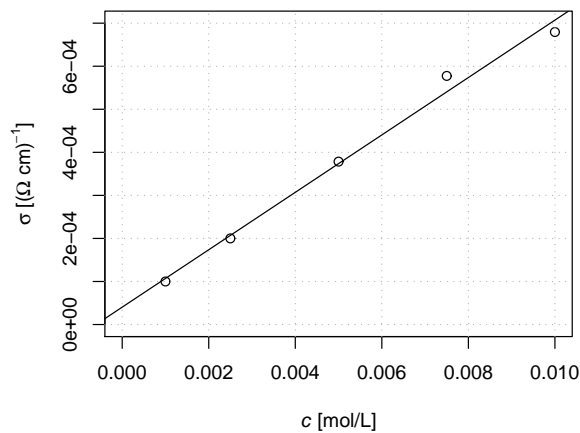}
  \caption{Linear dependence between the electrolyte concentration and
    its conductivity in a dilute electrolyte.}
  \label{fig:electrolyte}
\end{figure}

The example shows a few features of R which we want to comment on:
\begin{itemize}
\item Since we only have five data points, the data were entered
  directly into the program rather than being read from an external
  data file.  The \texttt{c()} function is used to concatenate data
  into a vector.
\item R encourages programming with vectors.  In the expression
  \texttt{k/resistance}, each element of the resulting vector is
  computed as a reciprocal value of the element of the original
  vector, multiplied by \texttt{k}.
\item Inspired by the \TeX\ typesetting system, R offers a capable
  method of entering mathematical expressions into graph labels using
  the \texttt{expression()} function \cite{Murrell:2000}.
\item The \texttt{plot()} function is the generic function for
  plotting objects in R; here we used it to produce a scatterplot.
\item The \texttt{lm()} function is used to fit any of several linear
  models \cite{Chambers:1991b}; we used it to perform a simple
  bivariate regression.  R is not overly talkative, and \texttt{lm()}
  does not produce any output on screen, it merely creates the object
  \texttt{cond.fit}, which we can later manipulate at will.  In the
  example we used \texttt{abline(cond.fit)} to plot the regression
  line atop the data points.  If we want to print the coefficients, we
  can use \texttt{coef(cond.fit)}.
\end{itemize}

The complete sequence of commands can be saved into a script file,
\emph{e.g.,} \texttt{electrolyte.R}, and can be executed at any later
time using the \texttt{source()} function:
\begin{verbatim}
> source("electrolyte.R")
\end{verbatim}

\subsection{Case 2: Nonlinear dependence}

Not all dependencies are linear.  Sometimes, they can be linearized by
an appropriate transformation (\emph{e.g.,} logarithmic), which
reduces the task of finding the optimal fitted curve back to the
linear case.  With a computer at hand, however, this is not necessary.
In this example, we examine the time dependence of voltage in a
circuit with two capacitors (figure~\ref{fig:scheme}).  The student
first charges the capacitor $C_1$ and then monitors the voltage as
$C_1$ discharges.  The data, saved as in a CSV format:

\begin{figure}
  \centering
  \includegraphics[scale=0.8]{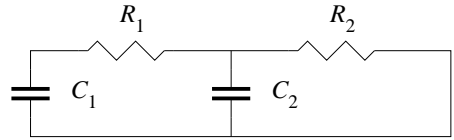}
  \caption{A circuit scheme with two capacitors.  In the experimental
    setup, $C_1 = C_2$ and $R_2 \gg R_1$.  $C_2$ is discharged at the
    beginning of the experiment.  The voltage $U$ on $C_1$ is recorded
    as a function of time $t$.}
  \label{fig:scheme}
\end{figure}

\begin{figure}
  \centering
  \includegraphics[angle=270,width=0.6\linewidth]{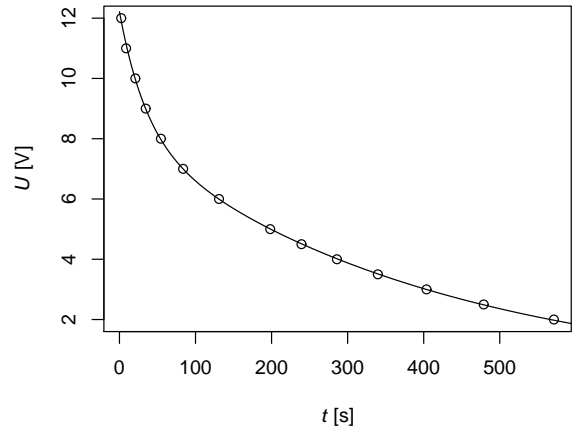}
  \caption{The voltage $U$ on the capacitor $C_1$
    (figure~\protect\ref{fig:scheme}) as a function of time $t$;
    measured data ($\circ$) and a fitted curve.}
  \label{fig:capacitor}
\end{figure}

\begin{verbatim}
"t [s]";"U [V]"
2,3;12,0
8,7;11,0
...
\end{verbatim}
The voltage $U(t)$ can be written as a sum of two exponentials:
\[ U(t) = A\, \mathrm{e}^{-t/\tau_1} + B\, \mathrm{e}^{-t/\tau_2} \; ,
\]
The task is to determine the two characteristic times, $\tau_1$ and
$\tau_2$.  The script below performs the task, and the resulting graph
is shown in figure~\ref{fig:capacitor}.
\begin{listing}{1}
C1 <- read.table("capacitor.csv", dec=",", sep=";", header = TRUE)

U <- C1$U..V.
t <- C1$t..s.

U.nls <- nls(U ~ A*exp(-t/t1) + B*exp(-t/t2),
           start = list(A = 5, B = 5, t1 = 10, t2 = 100))

plot(C1, xlab = expression(paste(italic(t)," [s]")),
     ylab = expression(paste(italic(U)," [V]")))

lines(0:600, predict(U.nls, list(t = 0:600)))
\end{listing}

A few features of R used in the example need some comment:
\begin{itemize}
\item \texttt{read.table()} returns the object of a class
  \texttt{data.frame}; we may visualize it as a table.  Individual
  columns in the table are addressed as {\ttfamily{\itshape
      table}\/\${\itshape column}}.  Column names are constructed from
  the table header, with all illegal characters being replaced by
  dots.  For easier manipulation, two vectors, \texttt{t} and
  \texttt{U}, were created from the appropriate columns of the table.
\item The actual nonlinear fitting is performed by the \texttt{nls()}
  function, which returns an object of the \texttt{nls} class.  Apart
  from the formula, we also supplied the initial guesses for the
  fitting parameters.  The values of the coefficients can be extracted
  by \texttt{coef(U.nls)}, and \texttt{summary(U.nls)} produces an
  even more informative report.
\item An object of an \texttt{nls} class also provides methods for the
  \texttt{predict()} function, which we used to plot a piecewise
  linear curve, which appears as a smooth curve due to the small step
  used.
\end{itemize}

\subsection{Case 3: Histogram}

\begin{figure}
  \centering
  \includegraphics[angle=270,width=0.6\linewidth]{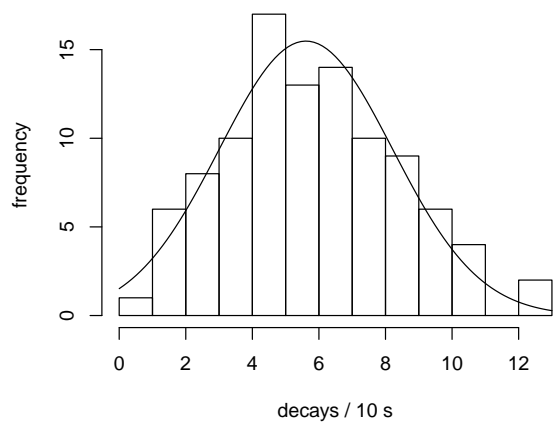}
  \caption{Experimentally obtained histogram of radioactive decay and
    a normal distribution ($\overline{x} = 5.61$, $\sigma_x = 2.60$)
    plotted on top of it.}
  \label{fig:radioaktiv}
\end{figure}

Our third example examines the statistics of radioactive decay.  Using
a Geiger counter, the student is required to record the number of
decays detected in a period of time (in our case, 10~s), and repeat
the measurement 100 times.  After subtracting the base activity, the
student computes the mean value $\overline{x}$, the standard deviation
$\sigma_x$, and plots a histogram.  The script shown below
demonstrates computing and plotting the histogram.  The final result
is shown in figure~\ref{fig:radioaktiv}.

\begin{listing}{1}
# detected decays per 10 s
x <- scan("decay.txt")

# subtract the base (normalized per 10 s)
x <- x - 2.68

avg <- mean(x)
stdev <- sd(x)

# draw the histogram
hist(x, breaks = seq(floor(min(x)), ceiling(max(x))),
     main = "", xlab = "decays / 10 s", ylab = "frequency")

# overlay the probability density function
curve(length(x)*dnorm(x, mean = avg, sd = stdev), add = TRUE)
\end{listing}

Again, a few comments.
\begin{itemize}
\item Since the format of the input data file is very simple (one
  figure -- the number of detected decays per 10~s -- per line), we
  have used the \texttt{scan()} function instead of
  \texttt{read.table()}; \texttt{x} is thus a vector rather than a
  data frame.
\item The histogram is plotted using the \texttt{hist()} function.
  The \texttt{breaks} option is used to specify bin size and
  boundaries.  By default, R uses Sturges' formula for distributing
  $n$ samples into $k$ bins:
  \[ k = \lceil \log_2 n + 1 \rceil \; . \]
\item The normal probability distribution function \texttt{dnorm()} is
  plotted over the histogram with the \texttt{curve()} function and
  \texttt{add = TRUE} option.
\end{itemize}

\subsection{Exporting the graph}

Finally, we need to make the graph available to other programs.  If we
want to include it in a paper or a report, this often means exporting
the graph as a PostScript file.
\begin{verbatim}
> postscript(file = "decay.eps", width = 5, height = 4.5)
> source("decay.R")
> dev.off()
\end{verbatim}
The options \texttt{width} and \texttt{height} specify its width and
height in inches.  Other useful output formats include PDF (using
\texttt{pdf()} function) and various bitmap formats (\texttt{bmp()},
\texttt{jpeg()}, \texttt{png()}, and \texttt{tiff()}).

\section{Discussion}

Having seen a few examples, we want to examine where R stands against
its competition, \emph{i.e.,} manual analysis and graphing,
spreadsheet software and commercial scientific graphing and/or
statistical software suites.

Producing graphs manually involves some processing of data with a
pocket calculator, determining the minimal and the maximal values of
data in $x$- and $y$ direction, choosing the appropriate scales on the
$x$- and $y$-axis, plotting the data points, drawing the best-fit line
and possibly using it in the subsequent analysis.  In the case of
histogram plotting, the student has to choose the bin size and manually
arrange the samples into bins.  It is instructive to perform this
whole process once or possibly a few times in order to get familiar
with all the steps required.  Since the amount of information a
physicist has to cope with often exceeds the human capacity of
processing in a tabular form and requires some visualization
technique, a certain degree of graphic, or visual literacy is
required.  Therefore, a lot of emphasis is given to teaching this
skill in the course of a physics curriculum
\cite{McDermott:1987,Beichner:1994,KarizMerhar:2009}.  However, we can
not overlook the fact that the processes involved in a manual
production of a graph are time-consuming, error-prone, and do not
contribute significantly to a better understanding of the problem
studied.  In this respect, we have to agree with earlier studies
\cite{Barton:1998} that if the purpose of the graph is data analysis
rather than acquiring the skill of plotting a graph, the time spent
with manual graph plotting is better spent discussing its meaning.

By employing the ``variable equals spreadsheet cell'' paradigm where
users can see their variables and their contents on screen,
spreadsheets have been extremely successful in bringing data
processing power to the profile of users who would have otherwise not
embraced a traditional approach to computer programming.  The
usefulness spreadsheets possess for teaching physics
\cite{Dory:1988,Webb:1993,Cooke:1997}, including laboratory courses
\cite{Guglielmino:1989,Krieger:1990}, has been recognized early on,
and they continue to find new uses (see, \emph{e.g.,}
\cite{Levesque:2008}).  However, the approach taken by spreadsheets
also has its drawbacks.  By placing emphasis on visualizing the data
itself, the relations between data are less emphasized than in
traditional computer programs, and it is generally more difficult to
debug a spreadsheet than a traditional computer program of comparable
complexity.  The fact that the variables are referred to by their grid
address rather than by some more meaningful name is an additional
hindrance factor.  Another sort of criticism concerns the graphical
output of spreadsheet programs.  While a skilled user can produce
clear, precise and efficient graphs with Excel (see, \emph{e.g.,}
\cite{Vidmar:2007} and the references therein), it is perceived
\cite{Su:2008} that less skilled users are easily misled into
excessive use of various embellishments (dubbed as ``chartjunk'',
\cite{Tufte:2001}) that do not add to the information content.
Telling good graphs from bad ones is not merely a subject of
aesthetics, there exists an extensive body of work in this area, see,
\emph{e.g.,}
\cite{Tukey:1977,Chambers:1983,Tufte:2001,Cleveland:1993,Cleveland:1994},
and \cite{Wilkinson:2005}.

While it is difficult to estimate the extent of its use in student
laboratories throughout the world, we can observe that R receives an
explosive growth of use in research laboratories, where one would
actually expect slower adoption due to the additional competition it
faces both from the commercial scientific graphing software and from
the commercial statistical software suites.  Thomson Reuters ISI Web
of Science\textsuperscript{\textregistered} shows that the R manual
\cite{R:Manual} has accumulated over 9000 citations since 1999, over
3000 of these in the year 2009 alone.  Interestingly, however, fewer
than 1\% of this total figure comes from physics, and of these, the
majority come from cross-disciplinary branches like geophysics and
biophysics.  

Using R brings considerable advantages.  It is free; students can
install a copy at home.  Using R, well-designed publication-quality
plots can be produced with ease.  It is an object-oriented matrix
language, which encourages thinking about programming problems on a
more abstract level.  It is an open source project; while its core is
actively maintained by a 20-strong core team which includes the
original author of S, over 1500 additional packages available on CRAN
were contributed by hundreds of people all over the world.  Last but
not least, by running scripts in batch mode rather than using R
interactively, reproducible research techniques \cite{Schwab:2000} are
encouraged. One further step in this direction is Sweave
\cite{Leisch:2003}, which allows creating integrated script/text
documents.

It also does not come without disadvantages.  Students entering the
laboratory course usually have no previous experience with R.  With
its complexity and a lack of point-and-click interface, the students
can not be expected to self-learn it; a quick introduction course has
been recommended even in the case of graduate students of
computational biology \cite{Eglen:2009}.  The best synergy might be
achieved when students take an introductory course in statistics
before the physics laboratory course or simultaneously with it.

In many ways, the position of R in the area of data analysis and
scientific graphing can be compared to the position \TeX\ holds in
typesetting texts with mathematical content.  Both are free software
packages, maintained by a tightly knit network of users/developers
rather than a corporation.  Both were designed as complete programming
languages, which enabled the community to extend their usefulness with
numerous add-on packages.  Both, admittedly, have a relatively steep
learning curve.  Both found their use in their specific niches, which
received a certain degree of neglect by the mainstream software
industry, and seem fairly entrenched there.

\section{Conclusions}

The extent of utilization that R has recently experienced in various
scientific disciplines signifies it is more than a marginal
phenomenon.  In the paper, we have demonstrated that R can be used
productively for for data analysis and graphing in an introductory
physics laboratory, and have illustrated its use on a few experiments
taken from an actual laboratory course.  The examples include a linear
dependence, a non-linear dependence, and a histogram.  The positive
and negative aspects of R were discussed against three options often
used for data analysis and graphing: manual graphing using grid paper,
general purpose spreadsheet software, and specialized scientific
graphing software.

\ack{This work has been supported by the Slovenian Research Agency
  through grant J3-2268.}

\section*{References}


\end{document}